%% file: main.tex
\documentclass[conference]{IEEEtran}

\usepackage{xspace}

\usepackage{balance}  
\usepackage{graphicx} 
\usepackage{times}    
\usepackage{url}      
\usepackage[utf8]{inputenc}
\usepackage[usenames, dvipsnames]{color}
\usepackage{balance}
\usepackage{moreverb}
\usepackage{amsmath}
\usepackage[utf8]{inputenc}
\usepackage{url}

\usepackage{listings}
\usepackage{color}
\usepackage{textcomp}
\usepackage{algpseudocode}
\usepackage{algorithm}
\usepackage{lmodern}
\usepackage{enumerate}
\usepackage{varwidth}
\usepackage{xcolor}
\usepackage{amssymb}

\let\emptyset\varnothing

\usepackage{multicol}

\definecolor{mygreen0}{rgb}{0, 0.75, 0}

\definecolor{myred1}{rgb}{1,0,0}
\definecolor{mygreen1}{rgb}{0, 1, 0}
\definecolor{myblue0}{rgb}{0, 0, 1}

\definecolor{myred2}{rgb}{1,0.5,0.5}
\definecolor{mygreen2}{rgb}{0.5, 1, 0.5}
\definecolor{myblue2}{rgb}{0.5, 0.5, 1}

\definecolor{mygreen}{rgb}{0, 0.25, 0}
\definecolor{myblue}{rgb}{0, 0, 0.75}
\definecolor{myred0}{rgb}{0.5,0,0}

\definecolor{listinggray}{gray}{0.98}
\definecolor{lbcolor}{rgb}{0.98,0.98,0.98}
\begin{document}
\date{}
\bibliographystyle{IEEEtran}
\newcommand{\projectName}{\emph{FIRM}\xspace}

\title{\LARGE \bf A \projectName Approach for Software-Defined Service Composition
}


\author{{\bf Pradeeban Kathiravelu} \\
{\small INESC-ID Lisboa} \\
{\small Instituto Superior Técnico,}\\
{\small Universidade de Lisboa}\\
{\small Lisbon, Portugal} \\
{\small pradeeban.kathiravelu@tecnico.ulisboa.pt}\\
\and 
{\bf Tihana Galinac Grbac} \\
{\small Faculty of Engineering} \\
{\small University of Rijeka}\\
{\small Rijeka, Croatia} \\
{\small tihana.galinac@riteh.hr}\\
\and 
{\bf Lu{\'\i}s Veiga} \\
{\small INESC-ID Lisboa} \\
{\small Instituto Superior Técnico,} \\
{\small Universidade de Lisboa}\\
{\small Lisbon, Portugal} \\
{\small luis.veiga@inesc-id.pt}}
\maketitle

\maketitle


\subsection*{Abstract}
Service composition is an aggregate of services often leveraged to automate the enterprise business processes. While Service Oriented Architecture (SOA) has been a forefront of service composition, services can be realized as efficient distributed and parallel constructs such as MapReduce, which are not typically exploited in service composition. With the advent of Software-Defined Networking (SDN), global view and control of the entire network is made available to the networking controller, which can further be leveraged in application level. This paper presents \projectName, an approach for Software-Defined Service Composition by leveraging SDN and MapReduce. \projectName comprises Find, Invoke, Return, and Manage, as the core procedures in achieving a QoS-Aware Service Composition.

\balance

\input{Introduction}

\input{related_work}	

\input{Architecture}

\input{Implementation}

\input{Evaluation}

\input{conclusion}

\tiny{\bibliography{references}}

\end{document}

%% file: Introduction.tex
\section{Introduction}
\label{sec:intro}
Complex enterprise services such as weather forecast, disaster prediction, and extraterrestrial activity monitoring systems~\cite{anderson2002seti} are traditionally deployed in clusters of servers distributed across the globe. Many systems consume other public services and mashup the information to produce a complex service composition. Service composition~\cite{rao2005survey} allows complex web services to be designed by composing simpler web services, and aggregating them to offer a complex execution of a business process or an enterprise requirement. As service composition involves execution of interdependent web services, multiple services can be composed such that they can be alternatives in offering the same service composition. 

SDN separates the control plane that controls the network from the data plane that consists of the switches that actually forward the network traffic~\cite{nunes2014survey}. Software-Defined Cloud Networking (SDCN) enables effective configuration of cloud deployments, extending the SDN paradigm to cloud-scale, with multiple heterogeneous physical entities and logical components, such as data centers, storage, and middleboxes~\cite{jain2013network}. Context-aware service composition has been proposed by leveraging SDN for the deployment of service composition~\cite{paganelli2014context}.

By deploying the service composition over an SDN, the controller is given an overall view of the service composition such as the service instances and the actual nodes that the services are hosted in. Further, web services engines contain the service health statistics on how many service requests were fulfilled by the system, and how many are on the fly. By making these statistics available to the controller, the service composition network can be made partially reconfigurable by the controller. 

MapReduce~\cite{dean2008mapreduce} is a programming model that can be executed in a parallel and distributed manner. While web services engines such as Apache Axis2~\cite{perera2006axis2} and Apache CXF~\cite{balani2009apache} are traditionally used for creating and hosting web services, MapReduce frameworks such as Apache Hadoop~\cite{white2012hadoop} can provide an effective distributed and scalable alternative for web services in service compositions, by realizing and offering the service implementations as MapReduce applications.

Effective scheduling and routing for the MapReduce traffic can be ensured by leveraging SDN. Moreover, congestion and failure of the underlying computing nodes and links can be monitored such that malfunctioning nodes can be dynamically removed or demoted from the computing cluster. Further, the execution can be replicated in an alternative node, or the routing of the MapReduce flows can be sent through an alternative route, when the network is observed to be severed partially in a few links or nodes.

Quality of Service (QoS) is crucial in enterprise service composition frameworks~\cite{charfi2007qos}. A service composition can be made QoS-aware, by considering the network congestion and load on the chosen web services or links, in choosing the web service deployments among the multiple available alternatives for the composition. Software-Defined Service Composition leverages the centralized global view of the network offered by the SDN controller in achieving the QoS-awareness. 

\projectName is an approach for large-scale QoS-aware service composition, leveraging SDN and approaches and paradigms such as MapReduce and dynamic programming. It proposes four procedures, named Find, Invoke, Return, and Manage for an application-aware service composition. Find procedure finds the appropriate service installations as the core services in the composition. Invoke invokes the chosen service deployments in an efficient and distributed environment. Return returns the results of the service compositions back to the user. Manage manages and orchestrates the service composition through a web service registry as well as the SDN controller.


Use of SDN and the adaptive service composition is found to be more beneficial for time consuming workflows such as MapReduce on big data, than the regular web services which are shorter in execution time. Exploiting the history information of the MapReduce execution cluster readily available to Hadoop, web services health status available to the web services registry, and a global view of the network available to the SDN controller, \projectName attempts to offer a QoS-Aware Software-Defined Service Composition.

In the upcoming sections, we will further analyze the proposed FIRM approach. Section II will address background information and related work on SDN and service composition. Section III discusses the \projectName approach for Software-Defined Service Composition and the design of the solution architecture implementing \projectName. Section IV further elaborates the prototype implementation details. Preliminary evaluations on \projectName are discussed in Section V. Finally, Section VI concludes the paper with the current state of the research and future work.

%% file: related_work.tex
\section{Background and Related Work}
\label{sec:related_work}

As the scale and complexity of the networks and systems is growing larger and larger with time, programmability of clouds and networked systems is researched intensively~\cite{campbell1999survey}. SDN facilitates effective management of large networked systems of cloud scale, increasing the reusability of the architecture and configurations, by providing a logically centralized control plane separated from the data plane that forwards the data~\cite{mckeown2009software}.


\subsection{Service Composition}
While Service Oriented Architecture (SOA) has been the forefront of the service composition~\cite{papazoglou2003service}, it is not uncommon to develop business processes and service compositions through other architectural paradigms such as Resource Oriented Architecture (ROA)~\cite{xu2008resource}. While alternative approaches pose their own implementation challenges and limitations, solutions based on parallel and distributed frameworks such as MapReduce and Dryad~\cite{isard2007dryad} to replace traditional service compositions can be more efficient and scalable at each service level. 

Web service registry plays a major role in QoS-aware service composition, as it offers a managed list of descriptions of the services. It stores the service end point descriptions and offers management and governance capabilities for web services. Many specifications have defined and standardized the service registry. Universal Description, Discovery, and Integration (UDDI)~\cite{curbera2002unraveling} offers a standardized directory structure as a registry of web services description. Distributed and effective service registries are built to minimize the load on the registry, to avoid registry being a single point of failure. Ad-UDDI is a distributed service registry, with an active monitoring mechanism~\cite{du2006ad}.


\subsection{Software-Defined Systems}
SDN controller manages the routing and forwarding rules, and updates the data plane which actually carries out the forwarding rules decided by the control plane~\cite{yeganeh2013scalability}. Data plane consisting of multiple instances can be centrally managed~\cite{kim2013improving}. OpenFlow~\cite{mckeown2008openflow} protocol is considered a driving force behind SDN. OpenFlow enabled switches can compose a software-defined network, along with one of the OpenFlow SDN controller implementations. Controllers execute, orchestrate, and manage the SDN algorithms and architectures in a physical network consisting of switches and hosts, or a network emulated by a network emulator such as Mininet~\cite{lantz2010network}.

Many controllers, from research as well as industry, implement the OpenFlow protocol. OpenDaylight~\cite{medved2014opendaylight}, Floodlight~\cite{wallner2013sdn}, Ryu~\cite{ryu2013framework}, Beacon~\cite{erickson2013beacon}, Maestro~\cite{ng2010maestro}, and ONOS~\cite{berde2014onos} are commonly cited controllers that implement the OpenFlow standard~\cite{mckeown2008openflow} to enable SDN. Most of the service composition and web services frameworks are developed in Java, and hence, a controller developed in Java such as ONOS, Floodlight, and OpenDaylight can be advantageous in developing extensions for service compositions. As OpenDaylight is a platform supported by the Linux Foundation with the sponsors from major players in the networking industry, choosing OpenDaylight for building a framework for software-defined service composition can be advantageous for enterprise adaptation.

\subsection{Related Work}
The demand for offering more configurability to service composition has been on the rise. The Next Generation Service Overlay Network (NGSON) is a specification offering context-aware service compositions~\cite{john2013research}, leveraging SDN and Network Function Virtualization (NFV)~\cite{batalle2013implementation} for the orchestration of services management and composition. The standardization effort of NGSON has been the motivation for many related works, focusing on efficient resource utilization and achieving pervasive services~\cite{liao2012toward}. However, the existing implementations so far have not exploited the programmable networks offered by SDN effectively, to offer a Software-Defined Service Composition. 

Top-k automatic service composition~\cite{deng2014top} exploits MapReduce to compose parallel and effective service compositions, and evaluates the efficiency of the proposed solution on large-scale service sets. This proves that MapReduce can indeed be an efficient alternative for regular web services frameworks, in service composition. However, this work fail short in providing and leveraging the inherent web services registry that the web services frameworks offer, where such an information can further be used to compose QoS-aware service compositions with the effective use of SDN. Palantir~\cite{yu2014palantir} leverages SDN to optimize MapReduce performance with the network proximity data. 

While offering the freedom to use MapReduce and other distributed execution frameworks for service composition, the existing functionality such as the availability of an organized web services registry with the end point descriptions should be highlighted in the improved approaches. Moreover, SDN should be leveraged to consume the health information and status of the deployed services readily available from the web service engine and registry, as well as the network information available to the controller itself. Hence, Software-Defined Service Composition can be designed by leveraging SDN for service compositions with web service technologies, or alternative parallel and distributed execution frameworks to replace the traditional web services.

%% file: Architecture.tex
\section{Solution Architecture}
\label{sec:arch}
\projectName is an architecture and framework for Software-Defined Service Composition. The architecture is separated from the deployment, offering a loose coupling between the logic and implementation. \projectName defines service compositions loosely, as a compound of multiple execution of services, which can be traditional web services or individual execution of parallel execution frameworks such as Hadoop or Spark~\cite{zaharia2010spark}. Hence, throughput or efficiency of large batch jobs can be improved by decomposing them as smaller interdependent services or jobs, that can execute independently, and later be composed to provide the final result. \projectName effectively delegates the service registry and management functionalities to the SDN controller extensions, foreseeing a customizable and adaptive web service composition framework.

\begin{figure}[!ht]
	\begin{center}
		\resizebox{\columnwidth}{!}{
			\includegraphics[width=\textwidth]{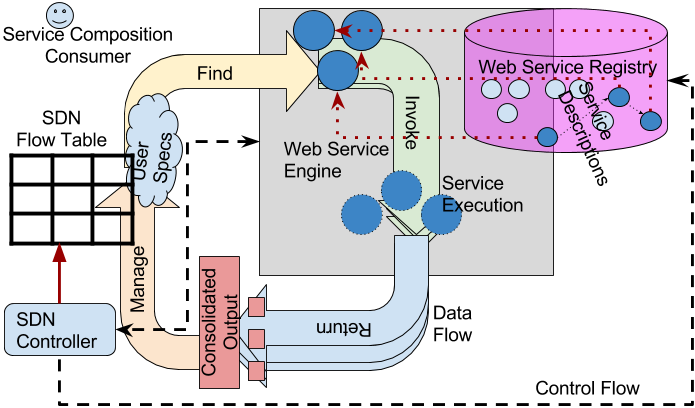}
		}
	\end{center}
	\caption{\projectName Procedures}
	\label{fig:firm}
\end{figure}

\projectName defines Find, Invoke, Return, and Manage as the core procedures, as shown by Figure~\ref{fig:firm}. The web service engine hosts the web services. Web service registry functions as a registry that stores the descriptions of web service deployments including the web service end points. Web service engines often dynamically store information such as the requests on the fly, requests completed, requests failed, for each of the services. Web service registry is connected to the web service engine to dynamically update the registry based on the changes in the service deployment and health status.


Upon receiving the service composition requests from the client, Find procedure identifies the services to be invoked by analyzing the alternative service end points from the web service registry. Invoke procedure consists of invoking each of the services identified for the service composition. Invoke procedure is distributed and executed in parallel at service granularity. Return procedure composes and returns the final output of the service composition request. Find procedure interacts mostly with the service registry, where Invoke and Return procedures interact with the web service engine. 

SDN controller, along with its \projectName service composition extensions, leverages the network topology and flow table information readily available to it, and also exploits the health information from web service engine. Manage procedure checks the health and congestion of the service deployments and reorganizes the services registry and controller extensions by prioritizing the service implementations dynamically. Hence \projectName offers a preferential list based on QoS parameters at network level as well as services or application level.

\subsection{\projectName Execution Flow}
\begin{figure}[!ht]
	\begin{center}
		\resizebox{\columnwidth}{!}{
			\includegraphics[width=\textwidth]{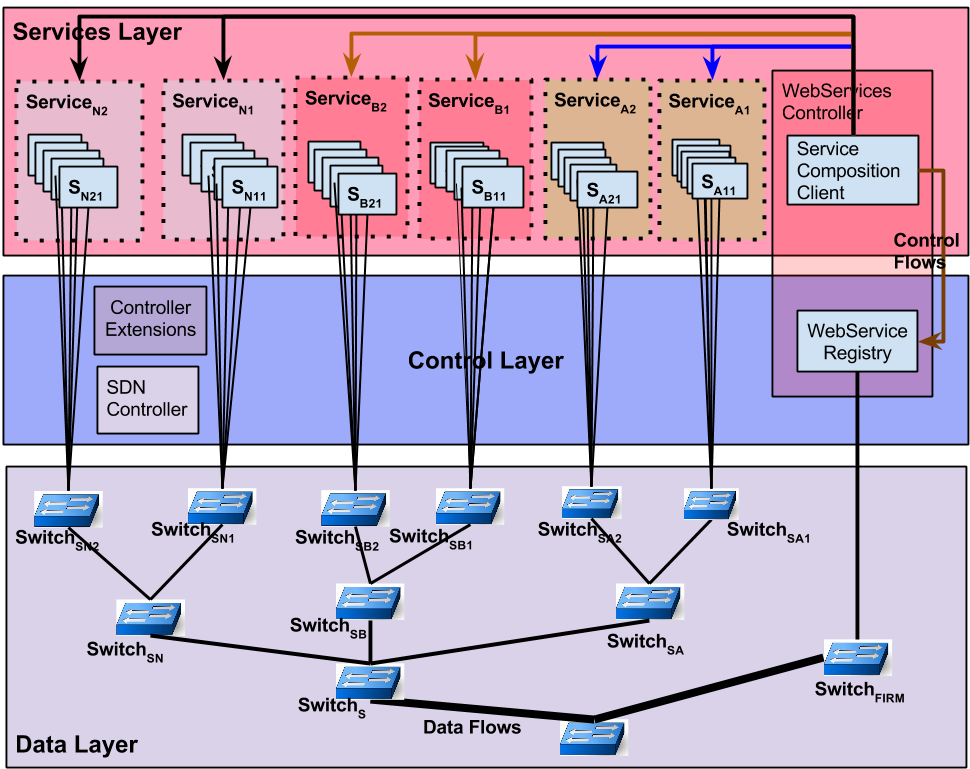}
		}
	\end{center}
	\caption{\projectName Deployment Architecture}
	\label{fig:depl}
\end{figure}

Figure~\ref{fig:depl} depicts the higher level deployment architecture of \projectName, with the switches organized in a fat tree topology~\cite{leiserson1985fat}. Data layer consists of the switches that are orchestrated by the controller in the control layer. The switches can also be organized in any other data center network topologies.

Here, \\
$\forall$ n $\in$ $\mathbb{Z}^{+}$; $\forall$ $\alpha$ $\in$ \{A, B, $\ldots$, N\}: $Service_{\alpha n}$ represents the $n^{th}$ implementation of $Service_{\alpha}$.

Similarly,
$\forall$ m $\in$ $\mathbb{Z}^{+}$: $S_{\alpha nm}$ represents the $m^{th}$ deployment of service implementation $Service_{\alpha n}$. Hence, for each service, $\sum\limits_{i=1}^n$ $m_{i}$  alternative deployments exists, with n different service implementations and a varying number $m_{i}$ of deployments for each implementation. Hence, a service composition consists of greater than or equal to min($\sum\limits_{i=1}^n$ $m_{i}$) number of alternative paths. Here each service in the composition can have $\sum\limits_{i=1}^n$ $m_{i}$ alternatives, and the service that has the minimum alternatives limits the number of potential alternatives for a service composition.


Control layer consists of the an SDN controller, controller northbound extensions for service compositions, and a web service registry. The controller extensions are core of the \projectName management, which effectively overlook the interaction between the SDN controller and web service layer. Controller extensions are either deployed along with the controller in the same node, or distributed to other instances, while still providing a unified logical view.

The web service registry holds a sorted list of services end points and service descriptions, with the input from the controller. \projectName registry is generalized to accommodate MapReduce services, in addition to the regular web services such as Axis2 or CXF. It is designed as a simple Java program holding the end points of the master node that receives the user requests, when the services are developed with MapReduce or other distributed execution frameworks. Services are deployed on the hosts that are connected to the switches in the network. In addition to the service hosts, the service composition client also is situated on the services layer.

Web services controller consists of both service composition client and web service registry. Service composition client receives the user requests and queries for the service deployment. It redirects the service calls to relevant service deployments, after consulting the service registry. At network level, the controller has the autonomy to decide one of the available alternative deployments for the same service implementation. Each service composition can be fulfilled by a series of alternative service implementations. 

\paragraph*{\textbf{Affinity and Stickiness in Service Invocations}}
Each service implementation has multiple deployments, offering scalability and load balancing to the service composition. Flow affinity is maintained such that once a service request is served by a deployment, further requests from the client are served from the same deployment by default. While this may seem as counter-intuitive from load balancing aspects, this avoids migration of state for the stateful invocations. 

\begin{figure}[!ht]
	\begin{center}
		\resizebox{\columnwidth}{!}{
			\includegraphics[width=\textwidth]{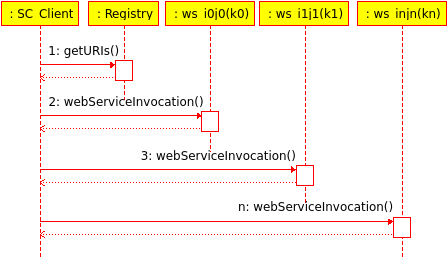}
		}
	\end{center}
	\caption{Service Composition with \projectName}
	\label{fig:init}
\end{figure}

Figure~\ref{fig:init} depicts a sample service composition workflow. Initially, the end point URIs (Uniform Resource Identifiers) of the potential services are retrieved from the registry by the service composition client. Once the web services to be invoked for the service composition are identified, they are executed by the service composition client. Results are returned to the service composition client by the respective services. If any service invocation depends on the results of a previous service invocation, the service blocks till the results are received. Services that do not have such dependencies are executed in parallel.

\subsection{\projectName Procedures}
The overall execution of \projectName is defined by Algorithm~\ref{alg:init} that calls the 4 major \projectName procedures.

\begin{algorithm}[!ht]
			\fontsize{9}{9}\selectfont			
	\caption{Execute \projectName}
	\label{alg:init}
	\begin{algorithmic}[1]
\Procedure{Execute()}{}
				\State 	 \colorbox{green!10}{$Manage()$}
\Repeat

		\For {	\colorbox{red!10}{($sc$ \textbf{in} $serviceCompositions$)}}
		
				\State 	 \colorbox{green!10}{endPoints $\gets$ $Find(sc)$}
		\For {	\colorbox{red!10}{($ep$ \textbf{in} $endPoints$)}}
				\State 	 \colorbox{green!10}{$Invoke(ep)$}
\EndFor
				\State 	 \colorbox{blue!10}{$Return(sc)$}
\EndFor
		\Until{\colorbox{red!10}{($aborted$)}}
		\EndProcedure
	\end{algorithmic}

\end{algorithm}

Manage procedure is invoked at the beginning of the execution at once, as shown by line 2, which is executed as an independent thread. Find, Invoke, and Return procedures are executed from the instance as a loop in the order, until \projectName is aborted. 

Service composition requests are processed in parallel by the framework, for each of the service composition as indicated in the lines 4 - 11. Each service composition is given as an input, which is further decomposed by the Find procedure to get the series of actual service end points, as in line 5. For each service end point that has been identified (line 6), Invoke procedure is invoked (line 7). Hence, Invoke procedure is invoked in parallel for multiple service end points. Finally, Return procedure consolidates the outputs of the service invocations and returns the final outcome of the service composition (line 9). 

Manage procedure is invoked once to initialize the system during the start up of \projectName, and waits to be triggered for updates after initialization execution in the beginning. This enables a learning behavior, as reportedly congested service end points and longer execution paths are avoided or `blacklisted' temporarily, as they are encountered and reported.

\subsubsection{\textbf{Manage}}
$Manage$ procedure is executed in parallel, independent of the service composition workflows which majorly consist of Find, Invoke, and Return procedure calls. $Manage$ consists of the initiating and maintenance flows interacting between the controller and service elements. Algorithm~\ref{alg:manage} depicts the Manage procedure.

\begin{algorithm}[!ht]
			\fontsize{9}{9}\selectfont			
	\caption{Managing the \projectName Framework}
	\label{alg:manage}
	\begin{algorithmic}[1]
\Procedure{Manage()}{}
				\State 	 \colorbox{green!10}{$initController()$}
				\State 	 \colorbox{green!10}{$initRegistry()$}

				\For {	\colorbox{red!10}{($servicEngine$ \textbf{in} $serviceEngines$)}}
				\State 	 \colorbox{green!10}{$init(serviceEngine)$}
				\State 	 \colorbox{green!10}{$updateRegistry(serviceEngine)$}
				\EndFor
\State 	 \colorbox{green!10}{$execPromoterThread(frequency)$}

		\Repeat

		\If {\colorbox{red!10}{($triggerUpdate(sc,serviceProperties)$)}}
				\For {	\colorbox{red!10}{($service$ \textbf{in} $serviceProperties$)}}
		\State \colorbox{green!10}{$updateFlowTable($}
		\par\hspace{80pt}\colorbox{green!10}{$service,serviceProperties)$} 
	
\EndFor
\Else
\State 	 \colorbox{green!10}{$sleep(frequency)$}

		\EndIf
				\Until{\colorbox{red!10}{($aborted$)}}

		\EndProcedure
	\end{algorithmic}

\end{algorithm}

At the beginning of the Manage procedure, the SDN controller  is initialized with the network. $initController()$ in line 2 initializes the controller and the extensions, with the flow table entries in the controller. $initRegistry$ invoked in line 3 initializes the web services registry with the information of the web service engines to be configured with \projectName. 

Each of the web service engine defined in the web service registry is initialized with the services deployed in them as shown by line 5. Initially the registry does not have any services information. As the web service engines are initialized, the registry is updated with the service information for each of the web service engines, by $updateRegistry()$ in line 6.


Manage procedure executes from a master instance, which is configured with the controller as an extension. Client threads from the service instances update the web services registry by invoking the master, when the preference order of the end point changes for the next invocation of web service composition. Manage procedure waits for the updates to be triggered while idling otherwise. 

Once the controller, registry, and the web service engines are initialized, $Manage$ procedure waits to be triggered for updates. An update is triggered when the service engine notices a delay in completing the web service invocation or when there are an increasing number of web service requests on the fly for a certain web service engine deployment. A reference to the triggering or offending service composition (sc), as well as the exact service properties (serviceProperties) consisting of the service, sorted list of preferred installation end points, and description are received along with the triggerUpdate method, as shown by line 10.

For each service that is included in the serviceProperties as the offending services, the SDN flow table is updated to reorder the preferred node to be used among the identical service deployments. updateFlowTable() invoked in line 12 updates the flow tables in the switches accordingly to route to the specific deployment of the web service implementation among the identical implementations.

\paragraph*{\textbf{Periodically Promoting the Demoted Deployments}}
As updateFlowTable() usually operates in demoting the service hosts from the routing tables, the services that are removed from the list must be added back periodically as they may have recovered from the overload, congestion, or the adverse status observed previously. execPromoterThread() invoked in line 8 ensures that the service deployments that have been demoted are periodically promoted back to serve the service invocations. This thread functions as a shuffling operation to ensure that no service is underused.

Algorithm~\ref{alg:promote} presents the promoter thread which executes independently in a timely manner to promote the `blacklisted' service deployments back to serve the web service requests. While this can be done by analyzing the health statistics of completed web service requests previously on the fly after `blacklisting' the service deployment, in simplistic approach it is often just decided based on a random event such as flipping a coin at random intervals, as shown by line 5 by invoking flipACoin() which returns a boolean value in a fair random manner. A frequency value is given as an input to the method to indicate how frequently the flipACoin() should be invoked.


\begin{algorithm}[!ht]
			\fontsize{9}{9}\selectfont			
	\caption{Promoter Thread}
	\label{alg:promote}
	\begin{algorithmic}[1]
\Procedure{ExecPromoterThread(frequency)}{}
		\Repeat
				\State 	 \colorbox{blue!10}{$promote.flag$ $\gets$ $false$}
								\State 	 \colorbox{green!10}{$sleep(frequency)$}
				\State 	 \colorbox{blue!10}{$promote.flag$ $\gets$ $flipACoin(frequency)$}
						\If {\colorbox{red!10}{($promote.flag$)}}

		\State 	 \colorbox{green!10}
		{{$promote.serviceID$ $\gets$ $readRandomLine($}} \par\hspace{80pt}\colorbox{green!10}{$listOfBlacklistedDeployments)$}

		\State \colorbox{green!10}{$promote(promote.serviceID)$} 
	
		\EndIf

		\Until{\colorbox{red!10}{($noSerivcesToPromote$)}}

		\EndProcedure
	\end{algorithmic}

\end{algorithm}


To avoid invoking the promote() too frequently, the promote flag is initialized to false (line 3), and the method sleeps for a given time for each iteration. At the specified frequency, the promote flag is reset by flipping a coin (line 5). If promote.flag value is set to true by the random event flipping coin, a random service deployment that was previously disabled is set to receive further web service requests in the future, till it is blacklisted again due to congestion or overload. The service to promoted is chosen by reading the list of black listed deployments at a random line. As each line represents a service, the service that is randomly chosen is promoted back to receive the web service invocations, as shown by line 8. As with the invocation frequency of promote(), choosing the service to be promoted can also be executed adhering to more intuitive algorithms than finding a random service. 

\subsubsection{\textbf{Find}}
$Find$ is the first procedure in the \projectName workflow, which finds the relevant service deployments to invoke for the service composition. $Find$ is depicted by Algorithm~\ref{alg:find}.

\begin{algorithm}[!ht]
			\fontsize{9}{9}\selectfont			
	\caption{Find Services}
	\label{alg:find}
	\begin{algorithmic}[1]
\Procedure{Find}{$sc$}

				\State 	 \colorbox{green!10}{$sc.serviceMap<services, properties>$ $\gets$ $parse(sc)$}
				\For {	\colorbox{red!10}{($service$ \textbf{in} $serviceMap.getKeys()$)}}
				\State 	 \colorbox{green!10}{$endPoints$ $\gets$ $getAvailableImpls(service)$}	
				\State \colorbox{blue!10}{$ep.properties$ $\gets$ $properties.get(service.getID())$}
				\State \colorbox{green!10}{$ep.service$ $\gets$ $getEndPoint(ep.properties)$}

				\State \colorbox{blue!10}{$sortedEndPoints.add(ep)$}
				
\EndFor
				\State \colorbox{blue!10}{\textit{\textbf{Return} $(sortedEndPoints)$}}

		\EndProcedure
	\end{algorithmic}

\end{algorithm}

Each service composition $sc$ is parsed into a map of services and properties, as in line 2. For each of the services in the service properties map, the list of implementations is derived from the registry (line 4). Service properties are retrieved from the service, as shown by line 5, retrieving information crucial for finding the service deployment best-fit for the composition. 

One of the service deployment end points, among the multiple potential service deployments is chosen, using the properties of the service as the parameter for the method getEndPoint() as in line 6. The end point is often a symbolic reference to the list of deployments of the same implementation. The exact deployment end point URI is chosen by the SDN extension, from the routing table as maintained by Manage procedure. Upon delegated to the network level, finding the exact host to invoke the service deployment is orthogonal to the Find procedure, as handled effectively by the controller.

The service end point as well as its properties are added to the sortedEndPoints variable, as shown in line 7. The service properties include service composition operators, which define how the service is related to the other services in the composition, whether the service execution can be distributed, should it block the subsequent service calls to the next service invocation, or can it be executed in parallel. Finally the list of sortedEndPoints is returned for the input of $sc$.

\paragraph*{\textbf{Minimizing Communication Overheads}} While the service deployment can be chosen following a naïve approach such as choosing the first in the list of available service end points, \projectName enables finding the service deployments that are in close proximity to each other, for each invocation of service composition. This minimizes the overhead caused by the inter-rack communication across the service deployments of the service composition, exploiting the network topology readily available to the controller.

\subsubsection{\textbf{Invoke}}
$Invoke$ is the second procedure in the \projectName workflow, which invokes one deployment for each of the service in the service composition. A service implementation is chosen by the Find procedure, and the controller is responsible for picking one of the exact service deployment end points for the execution. Manage procedure will be invoked if the status of a service deployment changes due to the service invocations. $Invoke$ procedure is depicted by Algorithm~\ref{alg:invoke}. 


\begin{algorithm}[!ht]
			\fontsize{9}{9}\selectfont			
	\caption{Invoke Services}
	\label{alg:invoke}
	\begin{algorithmic}[1]
	
\Procedure{Invoke}{$ep$}
\For {	\colorbox{red!10}{($ep'$ \textbf{in} $ep.properties.dependsOn()$)}}
\Repeat
			\State 	 \colorbox{green!10}{$sleep()$}
\Until {	\colorbox{red!10}{(ep'.out $\neq$ $\emptyset$)}}
\State 	 \colorbox{blue!10}{$ep.params.add(ep'.out)$}
\EndFor
			\State 	 \colorbox{green!10}{$ep.out \gets call(ep)$}
		\EndProcedure
	\end{algorithmic}

\end{algorithm}

Dynamic programming has been leveraged to effectively reuse the previously computed service execution results in the latter service executions or service compositions that depend on the previous. The service invocations that the current service depends on, are included into the service properties, such that they can be retrieved for the execution of the current service. The service invocation waits till the previous service executions that are the dependencies for the current execution are completed. 

Once the previous service execution results are available, they are added to the current service invocation parameters, as in line 6. Finally the current service call is initialized, and the result is stored as shown by line 8, for the future service executions that depend on the current.

\subsubsection{\textbf{Return}} Return procedure consolidates and returns the final result of the service composition execution back to the user. The web service engine updates its status on completed and pending web service requests. 


\subsection{Layered Architecture of \projectName}
\begin{figure*}[!ht]
  \includegraphics[width=\textwidth,height=10cm]{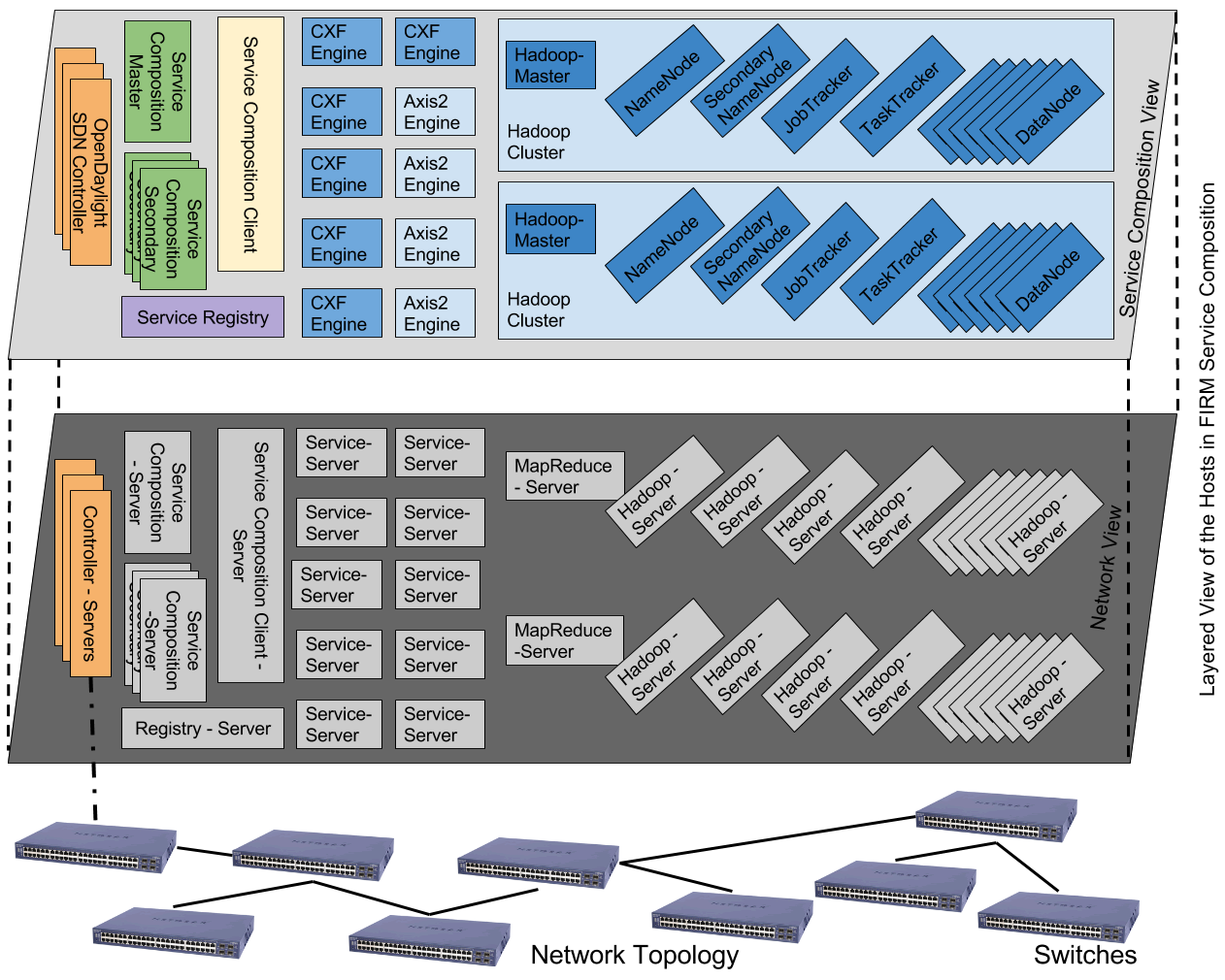}
	\vspace{-2em}
  \caption{Prototype Implementation Deployment - A Three-Dimensional View of \projectName Hosts and Topology}
  	\label{fig:layer}
\end{figure*}

\projectName has been developed following a layered architecture with a network view and a service composition view. Figure~\ref{fig:layer} shows the overall layered architecture of \projectName displaying both views. Network view consists of the hosts deployed on top of the network. Controller Servers are connected to the switches through OpenFlow protocol. All the servers are configured to form a network through the switches.

Service composition view looks into the same hosts through a higher level of abstraction. OpenDaylight, the default SDN Controller of \projectName, enables communication between the two views, as it is known from, and aware of, both views - through its southbound OpenFlow API to network view, and through its northbound user-facing APIs to service composition view. SDN controller is deployed as a cluster for scalability. Service Composition Master is responsible for the Manage procedure, which coordinates and manages the network in the service composition view. 

Along with the service composition master, multiple service composition secondary nodes are deployed to avoid single point of failure and overloading the master. Service registry consists of the description of the services hosted in the service engines. Service Composition Master gets the service health information from the web service engines and Hadoop Master instance of each of the Hadoop clusters and updates the SDN controller on the routing table. The service composition master uses service registry to get the list of services and service descriptions. It is further used to get the service end points which were earlier removed from the routing table by the Manage procedure back to the controller. Hence, the hosts hosting those services are added back to the flow tables after a specific time of blacklisting the congested or malfunctioning instances.

Service composition client is responsible for the Invoke procedure, where it simply invokes the relevant Axis2 or CXF services hosted in the Axis2 engine or CXF engine respectively. Moreover, it also connects to the Hadoop-Master instance which mimics the web service engines in producing the access point for the MapReduce cluster. In addition to the Hadoop-Master instance, each Hadoop cluster consists of a NameNode, JobTracker, TaskTracker, Secondary NameNode, and multiple DataNodes. The multiple DataNodes contribute to the Hadoop Distributed File System (HDFS). Service Composition client also returns the final outcome of the web service invocations to the service composition user.

%% file: Implementation.tex
\section{Implementation}
\label{sec:Implementation}
A prototype of \projectName was implemented to examine the feasibility of the proposed architecture with heterogeneous web service engines, Apache Hadoop 2.7.1, and OpenDaylight Lithium SDN controller. Apache Axis2 1.6.3 and Apache CXF 3.1.3 were used as the web service engines in the prototype developments. Service Registry was custom developed in order to accommodate descriptions of services deployed in heterogeneous web service engines, including the use of MapReduce frameworks such as Hadoop in the service composition workflow, replacing a traditional web service engine.

\subsection{Service Registry}
\projectName depends on the availability of multiple implementations of services offering same functionality, and multiple deployments of same implementations which can be chosen by the SDN controller extensions, where the service discovery is partially delegated to the network level. Highly congested service deployments, as observed from the service engine, are moved backwards or temporarily blacklisted from the controller extension. Hosts consisting of available service deployments that are high in the priority list are marked in the flow tables for the service composition workflows. 

An Nginx~\cite{reese2008nginx} style configuration was used for service descriptions in the registry. Given below is a sample service listing in the registry, with minimal description.

{\fontsize{9}{9}\selectfont				  
\begin{lstlisting}
services {
    service instance_count {
        type simple;
        impl axis2 {
            axa 192.168.0.104;
            ...
            axz 192.168.0.129;
        }    
        impl cxf {
            type jaxws_preliminary_ver {
                cxa 192.168.0.130;
            }
            type jaxws_ver_2 {
                update true;
                cxb 192.168.0.131;
            }
            type jaxrs {
                cxc 192.168.0.132;
            }
        }    
        impl mapreduce {
            mra 192.168.0.133;
            ...
            mry 192.168.0.157;
        }
    }
    service weather {
        type composition;
        entry_point 192.168.0.164;
        description {predicts the weather based on statistical models};
        services {
            instance_count {
            order 1;
            }
            adder {
            order 2;
            serialized false;
            }
            mean {
            order 3;
            serialized true;
            }
        }
    }    
}       
\end{lstlisting}
}

This format minimizes the text, and makes it easy to configure by the system administrators, as they are usually familiar with the Nginx-style configurations. Further service parameters can be added and parsed into the controller extension by extending the existing service registry reader APIs of \projectName.

Descriptions for the simple services define the service names, basic description, different installations, and multiple deployment end points for each of those installations. These alternative end points are basically multiple physical or virtual deployments of the same service code. Multiple service engines are deployed with the same service registry. Moreover, different implementations exist, even using the same service engine as depicted for CXF, where 3 implementations exist for the sample - 1 is a REST/JAX-RS based implementation, with the other 2 being SOA/JAX-WS based implementations. Further extended service specific parameters can be added, as shown by the property ``update'' for CXF implementation of jaxws\_preliminary\_ver. Service end point is the crucial information for simple web services.

Service compositions are defined by their underlying services. Execution order, whether the service can be distributed, should the service wait till the previous execution to complete, are a few of the regular properties included for the composition. An entry point is often defined for a service composition to compose and return the final output of the service composition in the Return phase. The entry point functions as the web service composition client. The registry can be modified through the configuration file at start up time, or can be modified manually or by \projectName dynamically, based on the preference of the services. Services' preference order is changed based on the load on the web services as observed by the service engine. 

If the client does not indicate a preference for specific service implementations and deployments in the service composition request, \projectName decides the service deployments to invoke in a QoS-aware manner. First, service implementations are often handled at the registry level, where service deployments for the same implementation is handled at SDN level. For the first web service request, controller extensions are invoked to find the potential services. Time taken to complete an individual service is measured at the web service engine, and the time taken to complete the service composition is measured at \projectName manager instance. The nodes hosting the web service deployments which consume more time to complete the service requests are moved downwards in the routing table to avoid further network flows to be routed to those nodes. Thus, finding the right service deployment is partially delegated to the SDN extension.

\subsection{Service Composition}
Requests to service compositions are described in the same way as service compositions are defined in the service registry. Users can execute complex queries of service compositions, by defining the service compositions with the services and service compositions defined in the registry.

Service compositions are given as tuples of services and the list of inputs for each of the services. The output of a service can be chained as the input of another, making the service composition. A simple GUI is also in place, implementing the relevant REST APIs, deploying the entire \projectName system as a web application. Given below is an example web service composition where the outputs of both service1 and service2 invocations are given as the input for the service3.

{\fontsize{9}{9}\selectfont				  
\begin{lstlisting}
<Service3,(<Service1, Input1>,<Service2, Input2>)>
\end{lstlisting}
}

Here Service1 and Service2 can be executed in parallel, while Service3 will have to wait till both Service1 and Service2 complete the execution, as it depends on their results as its input.

%% file: Evaluation.tex
\section{Evaluation}
\label{sec:Eval}
\projectName was evaluated for its performance and scalability in a combination of real and emulated service composition networks. Due to the limited accessibility to a 1024 host large scale multi-rack data center network, emulated networks were developed with \projectName, Mininet, and OpenDaylight, with web service frameworks and Hadoop. Performance of \projectName and Software-Defined Service Composition approach was benchmarked against the regular service compositions, for a complex service composition of weather prediction.

\subsection{Performance and Scalability}
Three major application scenarios were evaluated for the execution of multiple service composition requests: (i) Base service composition as offered by the web service engines and MapReduce frameworks; (ii) Service invocation with flow affinity improvements as proposed by \projectName, while not leveraging SDN and being completely agnostic to the network; (iii) With both affinity and congestion control offered by \projectName, leveraging SDN. As \projectName development and deployment followed an incremental approach, (ii) offered an intermediate state without deployment on SDN, while (iii) offered a complete implementation of \projectName and deployment on SDN. Approach (ii) finds the service deployments entirely at the service level using the service status information available to the web service engines.

\begin{figure}[!ht]
	\begin{center}
		\resizebox{\columnwidth}{!}{
			\includegraphics[width=\textwidth]{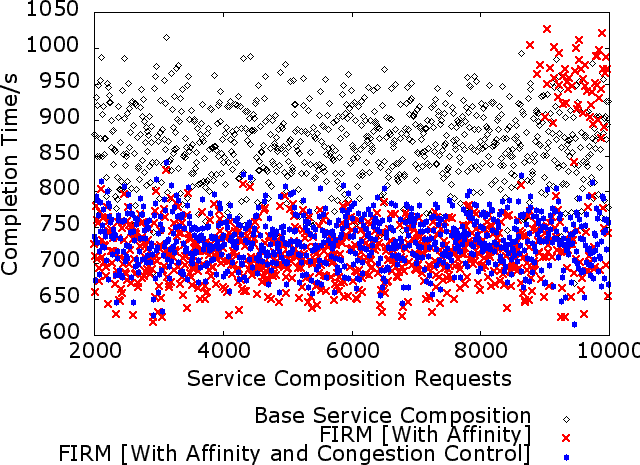}
		}
	\end{center}
	\caption{\projectName Approach With and Without SDN}
	\label{fig:eval}
\end{figure}

Figure~\ref{fig:eval} depicts the time taken for all these three scenarios. Base approach took up to 1000 seconds to complete the service compositions. The solution scaled well, efficiently across the large cluster of 1024 nodes. However, there was a considerable data transfer across the nodes and racks even for a single service composition, which increased the bandwidth consumption.

In the approach (ii), the network was observed to be adaptively scaling based on the service composition requests. However, since the network statistics and load was not monitored or considered, when the load of the web service requests went beyond a certain value, a few of the services faced overload and congestion. As a result, though the system scaled for reasonably large service compositions, it started to take much more time when the requests reached 9,000.

In the final congestion-aware deployment of \projectName, the nodes that were already serving larger number of services were avoided dynamically. While this initially seemed to consume more time than the second approach using the statistics available for the web service engine, this approach scaled uniformly even for much larger systems. 
\subsection{Load Balancing and Congestion Awareness}
\begin{figure}[!ht]
	\begin{center}
		\resizebox{\columnwidth}{!}{
			\includegraphics[width=\textwidth]{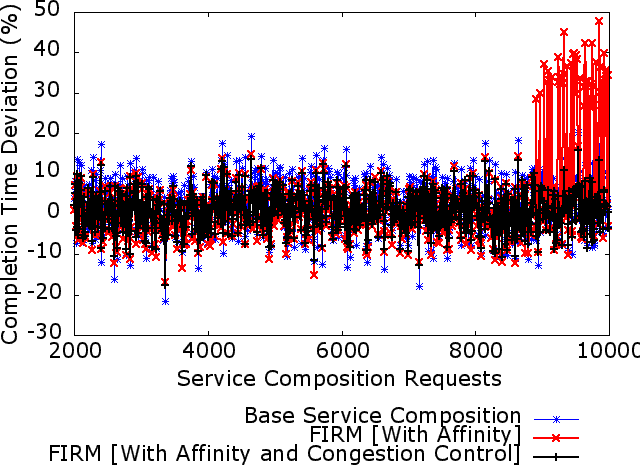}
		}
	\end{center}
	\caption{Deviation (\%) in the Completion Time}
	\label{fig:dev}
\end{figure}

Figure~\ref{fig:dev} depicts the deviation in completion time across multiple service composition executions, in order to find the imbalance in service distribution. The base approach handles the load balancing reasonably well, as Hadoop/MapReduce and web service engines are already optimized to distribute service requests across multiple instances. When attempting to distribute the load across multiple instances respecting flow affinity without leveraging SDN, there was a considerable overload in a few services. By leveraging SDN to dynamically route the traffic to the instances that are under-loaded, \projectName offers congestion awareness and balances the load even for higher concurrency levels effectively.

\begin{table}[!t]
\caption{Enhancements to Service Composition}
\label{table:enhancements}
\begin{tabular}{|c||c| |c|}
\hline
Feature & Existing Approaches & \projectName \\

\hline
Scalability & High & Higher   \\
Bandwidth Overhead & Low  & Low\\
Congestion Awareness & Low & High\\
Performance & High & Higher \\
Distributed Execution & High & Higher \\
\hline
\end{tabular}

\end{table}

Table~\ref{table:enhancements} compares the existing network agnostic approaches in service composition with SDN-based \projectName approach. While the related service composition frameworks offer a high scalability and performance, \projectName offers an even higher scalability and performance by leveraging SDN. Service executions should be executed with less communication overheads by avoiding frequent transfer of data and state across multiple servers, to minimize the bandwidth consumption. Both a network-agnostic and \projectName approaches produce low bandwidth overhead by offering flow affinity. While not compromising flow affinity, \projectName offers a more distributed approach to service composition with congestion awareness, which is lacking in the network agnostic service composition approaches. 


%% file: conclusion.tex
\section{Conclusion and Future Work}
\label{sec:conclusion}
\projectName proposes Software-Defined Service Composition by leveraging SDN for a QoS-aware service composition. Axis2 and CXF web service engines and Hadoop MapReduce framework are used for constructing the services for the compositions. Evaluation on the prototype implementation indicated that by utilizing the status information of the network available from SDN, service composition frameworks can be made scalable and efficient.

\projectName has been designed loosely coupled to specific technology, and avoids depending on web services as it can be implemented for any execution that can be made into a series of executions. As a future work, \projectName should be implemented for more distributed execution frameworks such as Dryad~\cite{isard2007dryad} and Apache Spark~\cite{zaharia2010spark} and evaluated for more service composition use case scenarios in real-world physical deployment environments.

While \projectName targets service composition with web services and distributed execution frameworks, Software-Defined Service Composition can be extended for the network functions, in service composition of middlebox actions, such as load balancing and firewall. There have been previous work on developing a framework for NFV~\cite{Palkar:2015:EFN:2815400.2815423}. Adopting Software-Defined Service Composition as an NFV framework for service function chaining should be benchmarked with the existing approaches.

\textit{\textbf{Acknowledgements:}} The work presented in this paper is supported by COST action 1304 Autonomous Control for a Reliable Internet of Services (ACROSS).


%% file: main.bbl
\begin{thebibliography}{10}
\providecommand{\url}[1]{#1}
\csname url@samestyle\endcsname
\providecommand{\newblock}{\relax}
\providecommand{\bibinfo}[2]{#2}
\providecommand{\BIBentrySTDinterwordspacing}{\spaceskip=0pt\relax}
\providecommand{\BIBentryALTinterwordstretchfactor}{4}
\providecommand{\BIBentryALTinterwordspacing}{\spaceskip=\fontdimen2\font plus
\BIBentryALTinterwordstretchfactor\fontdimen3\font minus
  \fontdimen4\font\relax}
\providecommand{\BIBforeignlanguage}[2]{{%
\expandafter\ifx\csname l@#1\endcsname\relax
\typeout{** WARNING: IEEEtran.bst: No hyphenation pattern has been}%
\typeout{** loaded for the language `#1'. Using the pattern for}%
\typeout{** the default language instead.}%
\else
\language=\csname l@#1\endcsname
\fi
#2}}
\providecommand{\BIBdecl}{\relax}
\BIBdecl

\bibitem{anderson2002seti}
D.~P. Anderson, J.~Cobb, E.~Korpela, M.~Lebofsky, and D.~Werthimer, ``Seti@
  home: an experiment in public-resource computing,'' \emph{Communications of
  the ACM}, vol.~45, no.~11, pp. 56--61, 2002.

\bibitem{rao2005survey}
J.~Rao and X.~Su, ``A survey of automated web service composition methods,'' in
  \emph{Semantic Web Services and Web Process Composition}.\hskip 1em plus
  0.5em minus 0.4em\relax Springer, 2005, pp. 43--54.

\bibitem{nunes2014survey}
B.~Nunes, M.~Mendonca, X.-N. Nguyen, K.~Obraczka, T.~Turletti \emph{et~al.},
  ``A survey of software-defined networking: Past, present, and future of
  programmable networks,'' \emph{Communications Surveys \& Tutorials, IEEE},
  vol.~16, no.~3, pp. 1617--1634, 2014.

\bibitem{jain2013network}
R.~Jain and S.~Paul, ``Network virtualization and software defined networking
  for cloud computing: a survey,'' \emph{Communications Magazine, IEEE},
  vol.~51, no.~11, pp. 24--31, 2013.

\bibitem{paganelli2014context}
F.~Paganelli, M.~Ulema, and B.~Martini, ``Context-aware service composition and
  delivery in ngsons over sdn,'' \emph{Communications Magazine, IEEE}, vol.~52,
  no.~8, pp. 97--105, 2014.

\bibitem{dean2008mapreduce}
J.~Dean and S.~Ghemawat, ``Mapreduce: simplified data processing on large
  clusters,'' \emph{Communications of the ACM}, vol.~51, no.~1, pp. 107--113,
  2008.

\bibitem{perera2006axis2}
S.~Perera, C.~Herath, J.~Ekanayake, E.~Chinthaka, A.~Ranabahu, D.~Jayasinghe,
  S.~Weerawarana, and G.~Daniels, ``Axis2, middleware for next generation web
  services,'' in \emph{Web Services, 2006. ICWS'06. International Conference
  on}.\hskip 1em plus 0.5em minus 0.4em\relax IEEE, 2006, pp. 833--840.

\bibitem{balani2009apache}
N.~Balani and R.~Hathi, \emph{Apache Cxf web service development: Develop and
  deploy SOAP and RESTful web services}.\hskip 1em plus 0.5em minus 0.4em\relax
  Packt Publishing Ltd, 2009.

\bibitem{white2012hadoop}
T.~White, \emph{Hadoop: The definitive guide}.\hskip 1em plus 0.5em minus
  0.4em\relax " O'Reilly Media, Inc.", 2012.

\bibitem{charfi2007qos}
A.~Charfi, R.~Khalaf, and N.~Mukhi, \emph{QoS-aware web service compositions
  using non-intrusive policy attachment to BPEL}.\hskip 1em plus 0.5em minus
  0.4em\relax Springer, 2007.

\bibitem{campbell1999survey}
A.~T. Campbell, H.~G. De~Meer, M.~E. Kounavis, K.~Miki, J.~B. Vicente, and
  D.~Villela, ``A survey of programmable networks,'' \emph{ACM SIGCOMM Computer
  Communication Review}, vol.~29, no.~2, pp. 7--23, 1999.

\bibitem{mckeown2009software}
N.~McKeown, ``Software-defined networking,'' \emph{INFOCOM keynote talk},
  vol.~17, no.~2, pp. 30--32, 2009.

\bibitem{papazoglou2003service}
M.~P. Papazoglou, ``Service-oriented computing: Concepts, characteristics and
  directions,'' in \emph{Web Information Systems Engineering, 2003. WISE 2003.
  Proceedings of the Fourth International Conference on}.\hskip 1em plus 0.5em
  minus 0.4em\relax IEEE, 2003, pp. 3--12.

\bibitem{xu2008resource}
X.~Xu, L.~Zhu, Y.~Liu, and M.~Staples, ``Resource-oriented architecture for
  business processes,'' in \emph{Software Engineering Conference, 2008.
  APSEC'08. 15th Asia-Pacific}.\hskip 1em plus 0.5em minus 0.4em\relax IEEE,
  2008, pp. 395--402.

\bibitem{isard2007dryad}
M.~Isard, M.~Budiu, Y.~Yu, A.~Birrell, and D.~Fetterly, ``Dryad: distributed
  data-parallel programs from sequential building blocks,'' in \emph{ACM SIGOPS
  Operating Systems Review}, vol.~41, no.~3.\hskip 1em plus 0.5em minus
  0.4em\relax ACM, 2007, pp. 59--72.

\bibitem{curbera2002unraveling}
F.~Curbera, M.~Duftler, R.~Khalaf, W.~Nagy, N.~Mukhi, and S.~Weerawarana,
  ``Unraveling the web services web: an introduction to soap, wsdl, and uddi,''
  \emph{IEEE Internet computing}, no.~2, pp. 86--93, 2002.

\bibitem{du2006ad}
Z.~Du, J.~Huai, and Y.~Liu, ``Ad-uddi: An active and distributed service
  registry,'' in \emph{Technologies for E-Services}.\hskip 1em plus 0.5em minus
  0.4em\relax Springer, 2006, pp. 58--71.

\bibitem{yeganeh2013scalability}
S.~H. Yeganeh, A.~Tootoonchian, and Y.~Ganjali, ``On scalability of
  software-defined networking,'' \emph{Communications Magazine, IEEE}, vol.~51,
  no.~2, pp. 136--141, 2013.

\bibitem{kim2013improving}
H.~Kim and N.~Feamster, ``Improving network management with software defined
  networking,'' \emph{Communications Magazine, IEEE}, vol.~51, no.~2, pp.
  114--119, 2013.

\bibitem{mckeown2008openflow}
N.~McKeown, T.~Anderson, H.~Balakrishnan, G.~Parulkar, L.~Peterson, J.~Rexford,
  S.~Shenker, and J.~Turner, ``Openflow: enabling innovation in campus
  networks,'' \emph{ACM SIGCOMM Computer Communication Review}, vol.~38, no.~2,
  pp. 69--74, 2008.

\bibitem{lantz2010network}
B.~Lantz, B.~Heller, and N.~McKeown, ``A network in a laptop: rapid prototyping
  for software-defined networks,'' in \emph{Proceedings of the 9th ACM SIGCOMM
  Workshop on Hot Topics in Networks}.\hskip 1em plus 0.5em minus 0.4em\relax
  ACM, 2010, p.~19.

\bibitem{medved2014opendaylight}
J.~Medved, R.~Varga, A.~Tkacik, and K.~Gray, ``Opendaylight: Towards a
  model-driven sdn controller architecture,'' in \emph{2014 IEEE 15th
  International Symposium on}.\hskip 1em plus 0.5em minus 0.4em\relax IEEE,
  2014, pp. 1--6.

\bibitem{wallner2013sdn}
R.~Wallner and R.~Cannistra, ``An sdn approach: quality of service using big
  switch’s floodlight open-source controller,'' \emph{Proceedings of the
  Asia-Pacific Advanced Network}, vol.~35, pp. 14--19, 2013.

\bibitem{ryu2013framework}
Ryu, ``Ryu sdn framework,'' 2013.

\bibitem{erickson2013beacon}
D.~Erickson, ``The beacon openflow controller,'' in \emph{Proceedings of the
  second ACM SIGCOMM workshop on Hot topics in software defined
  networking}.\hskip 1em plus 0.5em minus 0.4em\relax ACM, 2013, pp. 13--18.

\bibitem{ng2010maestro}
E.~Ng, ``Maestro: A system for scalable openflow control,'' TSEN
  Maestro-Technical Report TR10-08, Rice University, Tech. Rep., 2010.

\bibitem{berde2014onos}
P.~Berde, M.~Gerola, J.~Hart, Y.~Higuchi, M.~Kobayashi, T.~Koide, B.~Lantz,
  B.~O'Connor, P.~Radoslavov, W.~Snow \emph{et~al.}, ``Onos: towards an open,
  distributed sdn os,'' in \emph{Proceedings of the third workshop on Hot
  topics in software defined networking}.\hskip 1em plus 0.5em minus
  0.4em\relax ACM, 2014, pp. 1--6.

\bibitem{john2013research}
W.~John, K.~Pentikousis, G.~Agapiou, E.~Jacob, M.~Kind, A.~Manzalini, F.~Risso,
  D.~Staessens, R.~Steinert, and C.~Meirosu, ``Research directions in network
  service chaining,'' in \emph{Future Networks and Services (SDN4FNS), 2013
  IEEE SDN for}.\hskip 1em plus 0.5em minus 0.4em\relax IEEE, 2013, pp. 1--7.

\bibitem{batalle2013implementation}
J.~Batalle, J.~Ferrer~Riera, E.~Escalona, and J.~A. Garcia-Espin, ``On the
  implementation of nfv over an openflow infrastructure: Routing function
  virtualization,'' in \emph{Future Networks and Services (SDN4FNS), 2013 IEEE
  SDN for}.\hskip 1em plus 0.5em minus 0.4em\relax IEEE, 2013, pp. 1--6.

\bibitem{liao2012toward}
J.~Liao, J.~Wang, B.~Wu, and W.~Wu, ``Toward a multiplane framework of ngson: A
  required guideline to achieve pervasive services and efficient resource
  utilization,'' \emph{Communications Magazine, IEEE}, vol.~50, no.~1, pp.
  90--97, 2012.

\bibitem{deng2014top}
S.~Deng, L.~Huang, W.~Tan, and Z.~Wu, ``Top-automatic service composition: A
  parallel method for large-scale service sets,'' \emph{Automation Science and
  Engineering, IEEE Transactions on}, vol.~11, no.~3, pp. 891--905, 2014.

\bibitem{yu2014palantir}
Z.~Yu, M.~Li, X.~Yang, and X.~Li, ``Palantir: Reseizing network proximity in
  large-scale distributed computing frameworks using sdn,'' in \emph{Cloud
  Computing (CLOUD), 2014 IEEE 7th International Conference on}.\hskip 1em plus
  0.5em minus 0.4em\relax IEEE, 2014, pp. 440--447.

\bibitem{zaharia2010spark}
M.~Zaharia, M.~Chowdhury, M.~J. Franklin, S.~Shenker, and I.~Stoica, ``Spark:
  cluster computing with working sets,'' in \emph{Proceedings of the 2nd USENIX
  conference on Hot topics in cloud computing}, vol.~10, 2010, p.~10.

\bibitem{leiserson1985fat}
C.~E. Leiserson, ``Fat-trees: universal networks for hardware-efficient
  supercomputing,'' \emph{Computers, IEEE Transactions on}, vol. 100, no.~10,
  pp. 892--901, 1985.

\bibitem{reese2008nginx}
W.~Reese, ``Nginx: the high-performance web server and reverse proxy,''
  \emph{Linux Journal}, vol. 2008, no. 173, p.~2, 2008.

\bibitem{Palkar:2015:EFN:2815400.2815423}
\BIBentryALTinterwordspacing
S.~Palkar, C.~Lan, S.~Han, K.~Jang, A.~Panda, S.~Ratnasamy, L.~Rizzo, and
  S.~Shenker, ``E2: A framework for nfv applications,'' in \emph{Proceedings of
  the 25th Symposium on Operating Systems Principles}, ser. SOSP '15.\hskip 1em
  plus 0.5em minus 0.4em\relax New York, NY, USA: ACM, 2015, pp. 121--136.
  [Online]. Available: \url{http://doi.acm.org/10.1145/2815400.2815423}
\BIBentrySTDinterwordspacing

\end{thebibliography}
